\documentclass[preprint]{elsarticle}
\usepackage{graphicx}
\usepackage{array}
\usepackage{color}
\usepackage{amsmath}
\usepackage{SIunits}
\biboptions{sort&compress}

\begin{document}
\title{Thin interface limit for phase-field models of solidification with local mobility correction}

\author[Bochum]{Stephan Hubig\corref{cor}}
\ead{stephan.hubig@ruhr-uni-bochum.de}
\author[Bochum]{Raphael Schiedung}
\ead{raphael.schiedung@ruhr-uni-bochum.de}
\author[Bochum]{Ingo Steinbach}
\ead{ingo.steinbach@ruhr-uni-bochum.de}

\address[Bochum]{Interdisciplinary Centre for Advanced Materials Simulation, Ruhr-Universit\"at Bochum, Universit\"atsstr. 150, 44801 Bochum, Germany}
\cortext[cor]{Corresponding author}

\begin{abstract}

A new approach is developed to derive an analytical form for mobility corrections in phase-field models for pure material solidification.
Similar to the thin interface limit approach (Karma and Rappel, 1996) it seeks to remove systematic errors in the kinetics of phase-field models that arise due to the diffusivity of the interface. 
The new approach harnesses local information within the diffuse interface instead of an expansion within the interface peclet number $p_e=\eta/\delta$, the ratio of diffuse interface thickness to diffusion length.
Therefore it has no need for an approximation of $p_e$ being small. 
This results in a broader range of applicability of the approach for higher $p_e$.

\end{abstract}

\begin{keyword}
  model development \sep
  phase-field models 
\end{keyword}

\maketitle

\section{Introduction}
The phase-field method is established as the method of choice when it comes to the simulation of dendritic microstructure evolution  and a broad variety of phase transformations, with the prominent example of solidification processes \cite{Steinbach2013}.
Its central idea is to describe an interface between two phases implicitly as a transition region of finite thickness, in which an indicator function $\phi$, the phase-field variable, has a smooth transition between the phases.
The dynamics of phase transformation can then be described via a partial differential equation for $\phi$ and the property fields under consideration, here the temperature $T$.

When in 1994 the first dendrites were modeled with the phase-field method by Kobayashi \cite{Kobayashi1994a}, the method gained a significant prominence boost and was consequently investigated more thoroughly.
It was realized early on, that in the original simulations of Kobayashi the results depend strongly on the numerical discretization \cite{Merriman1994}, as well as on the spatial resolution, which is quantified by the interface peclet number $p_e$.
For quantitative numerical simulations, however, one needs a remedy for both types of errors.

Removing the first type of errors is an ongoing venture in research until today \cite{Glasner2001, Finel2018, Eiken2012} and requires special treatment of the numerical discretization.

To accomplish the latter one a correction of the systematic errors of the temperature field compared to the sharp interface solution is necessary.
This was first done by the so called sharp interface limit developed by Caginalp \cite{Caginalp1989}.
Later this approach was improved to the so called thin interface limit, which was established by Karma and Rappell using matched asymptotics in a first order expansion in the peclet number $p_e$ \cite{Karma1996, Karma1998a} and is the present-day state-of-the-art technique.

The local mobility correction approach presented here can be seen as an extension of the thin interface limits by providing a full analytic solution.
We also discuss effects of numerical discretization like interface locking.
This paper is organized as follows: First we present the theoretical derivation of the local mobility correction, and then test it in numerical simulations. To differentiate between numerical discretization errors and systematic errors due to finite interface width, first simulation results with constant driving force are examined and then a comparison between the new local mobility correction scheme and the established thin interface limit is done.

\section{Theoretical Background}
In this work a phase-field model, which describes the solidification of a pure material, is investigated.
It uses a double obstacle potential and a linearized driving force.
Its governing equations are written as
\begin{flalign}
\label{eq_kobayaschi_phi}
\dot{\phi} &= M \Bigl(   \sigma^* \bigl[ \Delta \phi + \frac{\pi^2}{\eta^2}(\phi - \frac{1}{2} )  \bigr]  + \frac{\pi}{\eta} \sqrt{\phi(1-\phi)} \Delta S ( T_m - T )  \Bigr) \\
\label{eq_kobayaschi_T}
\dot{T} &= \alpha \Delta T + k \dot{\phi}  ,
\end{flalign}
where $T$ is the temperature field and
$\phi$ is the phase-field variable, with $\phi = 0$ indicating presence of the liquid phase and $\phi = 1$ presence of the solid phase.

The phase-field mobility $M$ is going to be chosen either as a simple numerical parameter $M^\text{eff}$, or as a $\phi$-dependent function $M(\phi)$, depending on whether a thin interface limit or a local mobility correction scheme is employed.
$\sigma^*$ is the interface stiffness, $\eta$ the numerical interface width, $\Delta S$ is the entropy difference between bulk solid and liquid and  $T_m$ the melting temperature.
$\alpha$ is the thermal diffusivity and $k$ is the so called hypercooling temperature, which is related to the materials latent heat $L$, heat capacity $c_p$, and density $\rho$ via $k = L/(\rho c_p)$.

A phase-field model is regarded as quantitatively correct, if it reproduces the kinetic equation
\begin{align}
\label{eq_GibbsThomson}
v = M^\text{phys} ( \sigma^* \kappa + \Delta S ( T_m - T_\text{i} ) ),
\end{align}
of a sharp interface model.
Since in this work only the 1D-case is investigated, the interface curvature $\kappa$ will be set to zero.
The physical mobility $M^\text{phys}$ connects the thermodynamic driving force $\Delta S ( T_m - T_\text{i} )$, where $T_\text{i}$ is the interface temperature, to the interface velocity $v$.

In the thin interface limit approach by Karma and Rappel \cite{Karma1998a} formulas for
the effective interface mobility $M^\text{eff}$ are derived, which express it as a function of the desired physical mobility $M^\text{phys}$.
These formulas assure, that in the limit of vanishing peclet numbers $p_e = \eta/ \delta \to 0$  solutions of equation \eqref{eq_kobayaschi_phi} also fullfill equation \eqref{eq_GibbsThomson}.
Because of $\delta = \alpha/v$, the interface peclet number
\begin{align}
p_e = \frac{\eta v }{\alpha}
\end{align}
is also a measure for the interface velocity.

\begin{figure}
\centering
\includegraphics[width = 0.75\linewidth]{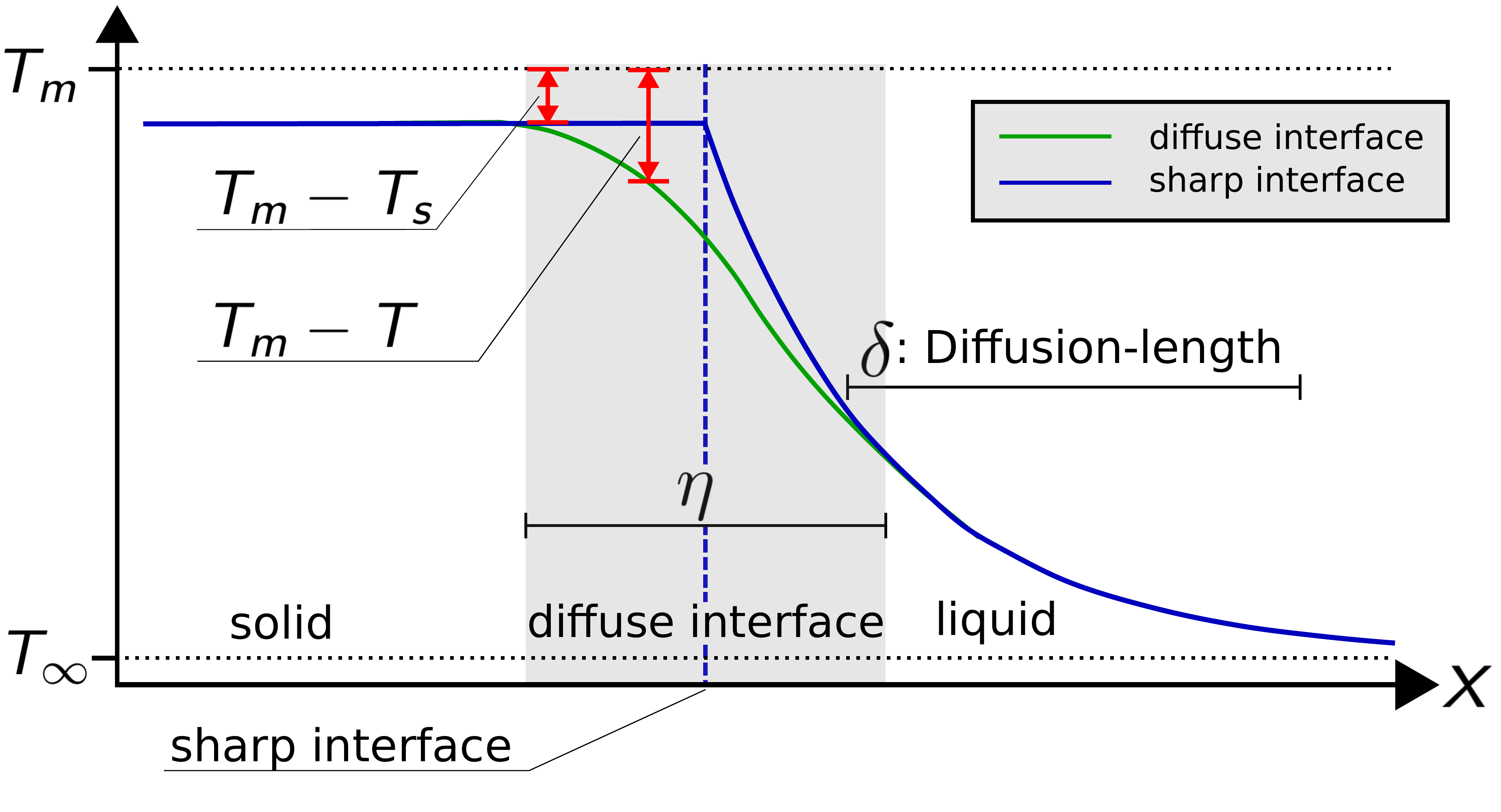}
\caption{ Diffuse and Sharp interface model in comparison. The red arrows indicate the deviation from equilibrium, which is varying over the interface.}
\label{fig_ThinSharpInt}
\end{figure}

To start let us consider two aspects, that make matching equation \eqref{eq_kobayaschi_phi} and \eqref{eq_GibbsThomson} challenging:
First of all, the driving force term in the phase-field model $\Delta S ( T_m - T )$ is locally varying with $T$, whereas in the sharp interface picture the driving force
$ \Delta S ( T_m - T_\text{i} )$ is a well defined single number.
Second, due to its locally varying nature, the driving force in the phase-field model has contributions originating from the bulk undercooling and not only the interfacial undercooling.
In the present case this would result in higher velocities than physically reasonable.
A sketch of this situation is given in figure~\ref{fig_ThinSharpInt}.
From this picture it also gets plausible, that in the limit $p_e \to 0$ variations in $T$ throughout the interface become small, and the drivingforce $\Delta S ( T_m - T )$  approaches a constant value throughout the interface. 
In the case of one dimensional steady state solidification, where the diffusion fluxes into the solid phase vanish, this driving force value will correspond to the temperature $T_s$ which is reached at the solid side of the diffuse interface.
Hence we identify $\Delta S ( T_m - T_s )$ as the physically reasonable driving force of the phase-field model.

In the sharp and thin interface approaches it is shown, that in the limit  $p_e \to 0$, where $T$ can be assumed to be constant, a solution of equation \eqref{eq_kobayaschi_phi} in one dimension is the traveling wave solution
\begin{align}
\label{eq_twsolution}
\phi _\text{tw}(x, t) = 
\begin{cases}
1   &\text{for } x < vt \\
\frac{1}{2} + \frac{1}{2}\cos ( \frac{\pi}{\eta}(x-vt)  )  &\text{for }  vt \leq x < vt + \eta \\
0   &\text{for } vt+\eta < x 
\end{cases}
\end{align}
with
\begin{align}
\label{eq_gibbth_rel}
v = M^\text{eff}  \Delta S ( T_m - T ).
\end{align}
For a constant temperature profile it is therefore shown, that by choosing $M^\text{eff} = M^\text{phys}$ the desired kinetics are reproduced.
If the temperature $T$ respectively the driving force  is not constant throughout the interface, neither $\phi_\text{tw}(x,t)$ will be a solution of the phase-field equation \eqref{eq_kobayaschi_phi}, nor will the kinetic relation \eqref{eq_GibbsThomson}  be reproduced correctly.
Similar to the aforementioned approach, our method also relies on the traveling wave solution, but aims to reproduce it in non approximative manner.

The main idea is now to define $M$ in equation \eqref{eq_kobayaschi_phi} as a  mobility function $M(\phi)$ in such a way, that the new phase-field equation  is still solved by $\phi_\text{tw}$ and the corresponding analogon to equation \eqref{eq_GibbsThomson} stays valid  even for locally varying $T$. 
This is possible, because $M(\phi)$ evaluates local information  via the phase-field variable and compensates local variations of $T$.
Therefore we call this approach local mobility correction.

In order to determine the analytical form of $M(\phi)$,
we first transform the governing equations \eqref{eq_kobayaschi_phi} and \eqref{eq_kobayaschi_T} in a moving frame with velocity $v$ via $z = x-vt $ and assume steady state evolution, so that we may replace $\frac{\text{d}}{\text{d}t}$ by $-v \frac{\partial}{\partial z}$.
Within the moving frame and inside the interface the traveling wave solution has the form $\phi_\text{tw} (z) = \frac{1}{2} + \frac{1}{2}\cos (\frac{\pi}{\eta}z) $. 
This enables us to first look for a form of the local mobility $M(z)$, which is dependent on the moving frame coordinate $z$ and not on $\phi$.
Since the traveling wave solution is monotonous, $M(z)$ can later be expressed as $M(\phi)$ by inverting $\phi_\text{tw} (z)$.
In the following the moving frame coordinate $z$ will be written as $x$.
 
Before we can determine the local mobility $M(x)$, we need to know the analytical form of $T(x)$, which is governed by equation \eqref{eq_kobayaschi_T}. Under the above mentioned conditions  equation \eqref{eq_kobayaschi_T} is
\begin{align}
-v \partial _x T = \alpha \partial_x^2 T  - vk \partial_x \phi_\text{tw}.
\end{align}

This ordinary differential equation  can be solved by first integrating directly with the initial conditions $T(0) = T_\text{s}$ and $T'(0) = 0$, which results in the equation
\begin{align}
\partial _x T 	&= \frac{v}{\alpha} \Bigl( T_s - T - k (1-\phi_\text{tw}) \Bigr)\\
	&= -  \frac{v}{\alpha} T 
	+  \frac{v}{\alpha} \Bigl( T_s - \frac{k}{2} \Big[1- \cos(\frac{\pi}{\eta} x)\Bigr] \Bigr).
\end{align}
Using the Ansatz $T(x) = C(x) \exp(-\frac{v}{\alpha}x)$ we arrive at
\begin{align}
\partial _x C(x) = \frac{v}{\alpha} \Bigl(  T_s - \frac{k}{2} \Bigl[ 1 - \cos(\frac{\pi}{\eta} x) \Bigr] \Bigr) \exp(-\frac{v}{\alpha}x).
\end{align}
Directly integrating this  leads to the temperature profile
\begin{align}
 T_R(x) =  T_\text{s} - v A_v(x),
\end{align}
where $A_v(x)$ is given by
\begin{align}
A_v(x) := \frac{k}{2v} \Bigl(  1- \frac{1}{\pi^2 \alpha^2 + \eta^2 v^2} \Bigl[  \pi^2 \alpha^2 \exp(-\frac{v}{\alpha}x) \\ + \pi \alpha v\eta \sin (\frac{\pi x}{\eta}) 
+\eta^2v^2 \cos (\frac{\pi x}{\eta})   \Bigr]    \Bigr)  . \notag
\end{align}
Note that $A_v$ if finite in the limit $v\to 0$: $\lim_{v \to 0} A_v(x) = \frac{k}{2\alpha}( x + \frac{\eta}{\pi} \sin(\frac{\pi x}{\eta}) )  $.

Inserting $\phi_\text{tw}(x)$ and $T_R(x)$ in equation \eqref{eq_kobayaschi_phi} results in
\begin{align}
\label{eq_pf_velocity}
-v \frac{ \partial \phi }{ \partial x } = - M(x) \frac{\partial \phi}{\partial x} \Delta S (T_m - T_R(x) ).
\end{align}
Straightforward rearranging delivers the equation 
\begin{align}
v 	=	\frac{M(x)}{ 1 -  M(x) \Delta S A_v(x)  }   \Delta S (T_m - T_\text{i}),
\end{align} 
where we identified $T_\text{s}$ as the interfacial temperature $T_\text{i}$ from the sharp interface picture.
The prefactor in front of the driving force $\Delta S (T_m - T_\text{i})$ is the physical interfacial mobility $M^\text{phys}$, which is reproduced by our phase-field model.
According to this by choosing
\begin{align}
\label{eq_local_correction}
M(x) = \frac{M^\text{phys}}{1+ M^\text{phys} \Delta S A_v(x) }
\end{align}
it is guaranteed, that the modified phase-field model reproduces the physical mobility $M^\text{phys}$.
Note that $A_v(x)$ is always positive, as long as $x>0$ holds  and this holds since $x \in [0,\eta]$ in the interfacial region.
This is reasonable since the locally corrected mobility should always be smaller than the physical one,
so that it can correct for the artificially high driving force within the interface.

In order to get a form of the corrected Mobility $M(\phi)$, which is then also valid in the original and not only the moving coordinate frame, and in order to make use of the local mobility correction in real simulations, the $x$-dependence of $M$ is now expressed by the local value of the phase-field variable $\phi$ itself.
The inversion of the traveling wave solution delivers
\begin{align}
\label{eq_phi_inversion}
 x = \frac{\eta}{\pi} \text{arcsin}( 2\phi - 1 ).
\end{align}
By inserting \eqref{eq_phi_inversion} into \eqref{eq_local_correction}, the $\phi$-dependent form $M(\phi)$ of the local mobility correction is easily obtained.

To derive a non local mobility correction in analogy to \cite{Karma1998a}, one can do the analysis above with $M$ defined as an effective numerical mobility $M^\text{eff}$. 
This results in the equation $v \frac{ \partial \phi }{ \partial x } =  M^\text{eff} \frac{\partial \phi}{\partial x} \Delta S (T_m - T_R(x) )$. Since this cannot be valid per point,
 we integrate by $\int_0^\eta \text{d}x$. 
Moreover we use the above mentioned limit $v\to 0$ for $A_v$, and than by rearranging and identifying the physical mobility as before, we get the equation
\begin{align}
\label{eq_nonlocal_correction}
M^\text{eff} = \frac{M^\text{phys}}{1+ M^\text{phys} \Delta S k \eta/(8\alpha) }.
\end{align}
The effective Mobility $M^\text{eff}$  corrects for the overestimated driving force like the local Mobility $M(x)$, but only in an average manner.

\section{Numerical implementation}
\label{sec_numerics}
\subsection{Numerical Methods}

In order to test the applicability  of local mobility corrections and examine possible benefits, one dimensional phase-field simulations are carried out, both with the local and the non local version of the mobility correction from equations \eqref{eq_local_correction} and \eqref{eq_nonlocal_correction} respectively.

For the numerical implementation of the non local mobility correction the space discretized version 
\begin{align}
\label{eq_method_local}
\dot{\phi}_i = M^\text{eff} \Bigl(   \sigma^* (  \frac{\phi_{i+1} + \phi_{i-1} -2\phi_{i} }{\Delta x^2} + \frac{\pi^2}{\eta^2}(\phi _i - \frac{1}{2} )  )  + \frac{\pi}{\eta} \sqrt{\phi _i(1-\phi _i)} \Delta S ( T_m - T_i )  \Bigr) 
\end{align}
of the phase-field evolution equation \eqref{eq_kobayaschi_phi}
with $M^\text{eff}$ chosen according to equation \eqref{eq_nonlocal_correction} is used.
For the numerical implementation of the local mobility correction 
\begin{align}
\label{eq_method_nonlocal}
\dot{\phi}_i = &M_0 \sigma^* \Bigl(      \frac{\phi_{i+1} + \phi_{i-1} -2\phi_{i} }{\Delta x^2} + \frac{\pi^2}{\eta^2}(\phi _i - \frac{1}{2} )   \Bigr) 
\\  &+  M(\phi_i) \Bigl( \frac{\pi}{\eta} \sqrt{\phi _i(1-\phi _i)} \Delta S ( T_m - T_i )  \Bigr) \notag
\end{align}
with $ M(\phi_i)$ chosen according to equation \eqref{eq_local_correction} is used.

In both cases the space discretized version of the heat conduction equation \eqref{eq_kobayaschi_T} is
\begin{align}
\dot{T}_i = \alpha \frac{ T_{i+1} +  T_{i-1} -2 T_{i} }{\Delta x^2} + k \dot{\phi}_i .
\end{align}

One important point is the introduction of a constant auxiliary mobility $M_0$ in equation~\eqref{eq_method_nonlocal}. 
This is done, because using the local mobility $M(\phi_i)$ in front of the stabilization operator $\Delta \phi - \frac{\pi^2}{\eta^2}(\phi-0.5)$ has shown to cause severe biases in the resulting velocities.
Since the stabilization operator pushes the numerical solution towards the traveling wave shape, it appears more appropriate to let it do so in an spatially uniform manner.
We will refer to  equation~\eqref{eq_method_nonlocal} as constant stabilization method.

To solve the time evolution an explicit euler scheme is applied: at time $t_m = m \Delta t$ the time derivative is approximated via $\dot \phi_i^m \approx (\phi_i^{m+1} - \phi_i^m )/\Delta t$, where $m$ and $m+1$ are indicating function values at timepoints $t_m$ and $t_{m+1}$.

The calculation of the velocity $v$, which needs to be known in order to calculate the local mobility $M(\phi_i^{m})$, is obtained by accessing information of the phase-field variable one time step earlier:
\begin{align}
v =   -\frac{\partial \phi}{\partial t} \Bigl(\frac{\partial \phi}{\partial x} \Bigr)^{-1}  = - \frac{2\Delta x}{\Delta t}(\phi_i^{m} - \phi_i^{m-1}) / (\phi_{i+1}^{m} - \phi_{i-1}^{m}).
\end{align}

\subsection{Choice of numerical parameters}
Model parameters as well as numerical discretization constants have to meet several conditions in order to grant stable simulations.
Because of the usage of en explicit time stepping scheme the stability criterion for the laplacians in the phase-field and the temperature field equation \cite{Press1991}
\begin{align}
\label{eq_numstab_temp}
\Delta t < \frac{\Delta x^2}{2\alpha} \\
\label{eq_numstab_phi}
\Delta t < \frac{\Delta x^2}{2\sigma^* M^\text{eff}}
\end{align}
 have to be fullfilled.
In order to avoid, that the interfacial region $0< \phi < 1$ gets spread by the varying driving forces beyond scales of $\eta$, the condition $\Delta G < \frac{\pi \sigma^*}{\eta}$
has proven to be useful.
To meet this condition we choose the interfacial energy, which has no physical meaning in the one dimensional problem, to be
\begin{align}
\label{eq_intf_stab}
\sigma^* =  \frac{\Delta S \Delta T_\text{i} \eta}{\pi \delta_\text{I}}
\end{align}
whith $\delta_\text{I} = 0.8$. $\Delta T_\text{i}=T_m - T_\text{i}$ is the driving force generating interface undercooling.

The smallest lengthscale, which has to be resolved in phase-field simulations, is the interface width $\eta$.
The number $n$ of gridpoints distributed across this length scale determines the spatial resolution of the simulation and is typically chosen to be $n = \frac{\eta}{\Delta x} \geq 5$.
In this work it is assumed, that, besides all material and process parameters ($\alpha$, $M^\text{phys}$, $\Delta S$, $T_m$, $\Delta T$), the interface width $\eta$ is given.
With knowing $\eta$ and the desired numerical spatial resolution $n$ the gridspacing $\Delta x$ can be easily calculated.
The timestep $\Delta t$ is then calculated according to equation~\eqref{eq_numstab_temp} via 
\begin{align}
\label{eq_numstab_temp_res}
\Delta t = \delta_s \frac{\Delta x^2}{\alpha} 
\end{align}
with $ \delta_s = 0.2 $
to assure numerical stability of the temperature solver.
Note that due to condition~\eqref{eq_intf_stab} this already implies numerical stability of the phase-field as long as $p_e \leq \pi \delta_\text{I}$, whereas for $p_e \geq \pi \delta_\text{I}$ numerical stability of the latter becomes the more restrictive condition.

The constant $M_0$ is chosen such, that in each timestep the dimensionless stabilization operator $\frac{\eta^2}{\Delta x^2} (\phi_{i+1} + \phi_{i-1} -2\phi_{i}) + \pi^2(\phi_i - \frac{1}{2}) $ contributes the highest possible fraction to the phase-field variable $\phi$ without causing numerical instabilities.
This can be achieved analogously to equation~\eqref{eq_numstab_temp_res} by setting
\begin{align}
\label{eq_M0choice}
M_0 = \delta_s \frac{\Delta x^2}{\Delta t \sigma^*}.
\end{align}

The used simulation parameters were: $T_m=1$, $k=0.1$, $\Delta T_\text{i} =0.05$, $\eta = 1$, $\sigma^* = 0.01$, $\Delta S=0.6$ and $\alpha = 100.0$.
The system was initialized by setting $\phi$ as a step function and setting the temperature field to a constant value of $T_m - \Delta T_\text{i}$ in the solid region and $T_m - \Delta T_\text{i} - k$ in the liquid region.

\section{Results and discussion}
\subsection{Simulations with constant driving force}
\begin{figure}
\centering
\includegraphics[width = 0.75\linewidth]{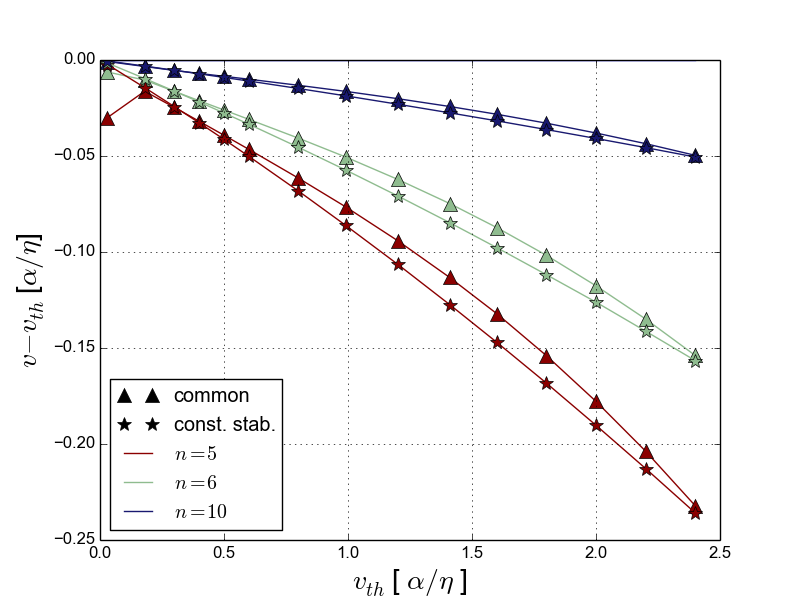}
\caption{ Results of constant driving force simulations. Different spatial resolutions $n=\eta/dx$ are indicated with colors, different numerical methods indicated with markers.}
\label{fig_constdrivforce}
\end{figure}

Since mobility correction schemes are meant to compensate for the varying driving force in the diffuse interface and not to correct the errors araising from numerical discretization, both types of errors have to be distinguished. 
For this purpose simulations with constant driving force were performed, in which only the equation
\begin{align}
\dot{\phi} &= M^\text{eff} \Bigl(   \sigma^* \bigl[ \Delta \phi + \frac{\pi^2}{\eta^2}(\phi - \frac{1}{2} )  \bigr]  + \frac{\pi}{\eta} \sqrt{\phi(1-\phi)} \Delta G  \Bigr) 
\end{align}
with constant $\Delta G := \Delta S \Delta  T_\text{i} $ was solved.
By varying the mobility $M^\text{phys}$, which is equal to $M^\text{eff}$ in the constant driving force case, the phase-field model is tweaked towards reproducing different theoretical velocities $v_\text{th}= M^\text{phys}\Delta G$.
In figure~\ref{fig_constdrivforce} the difference between actual and theoretical velocity $v-v_\text{th}$ is shown against $v_\text{th}$.
The velocities were measured by tracking the point with $\phi = 0.5$.
Those results show the known effect, that the numerical discretization is causing a decreased velocity  \cite{Merriman1994}.
It gets more severe for lower spatial resolutions and vanishes asymptotically for high spatial resolutions.

To get an idea about how the simulation kinetics are influenced by the constant stabilization method and by different weightings between stabilization operator and driving force term in normal spatial discretization, we compared the reproduced kinetics of both methods.

In the case of normal spatial discretization the driving force and mobility were weighted by a factor $f$ according to $\Delta G \to f \Delta G$ and $M^\text{eff} \to M^\text{eff}/f$. 
The theoretically expected velocity for different $f$-factors is the same.
For higher $f$ the stabilization operator has less weight relative to the driving force term.
Therefore we expect the simulated traveling wave profile to be more distorted.
It can be hypothesized that this will lead to errors also in the kinetics.

\begin{figure}
\centering
\includegraphics[width = 0.75\linewidth]{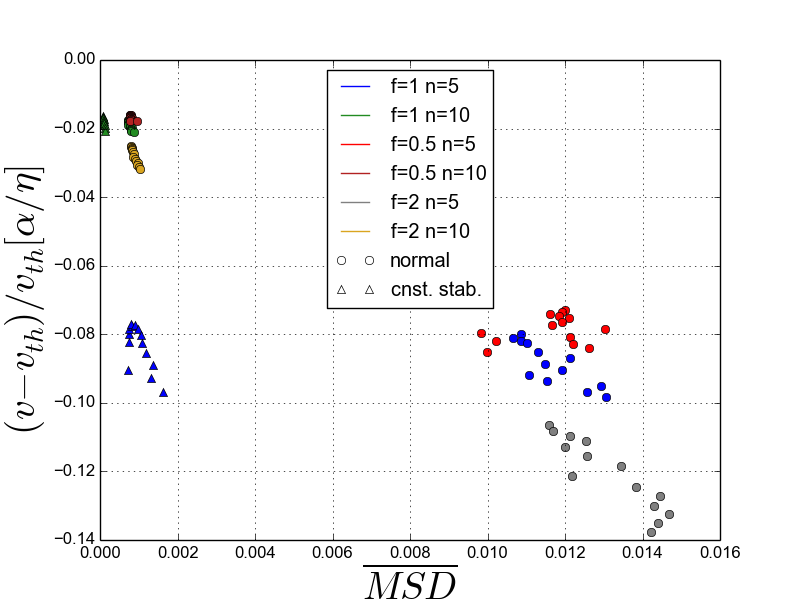}
\caption{ The correlation between time averaged mean squared deviation values $\overline{\text{MSD}}$ and the relative deviations of velocity from the theoretical velocity  $(v-v_{th})/v_{th}$ for different numerical methods, numerical resolutions $n$ and different weighting factors $f$. Note that the $f$-factor makes no difference for the constant stabilization operator. Those results are displayed in the colors for $f=1$.}
\label{fig_CorrPlot}
\end{figure}

To test for this hypothesis, the traveling wave profiles were analysed.
For this purpose the analytical form of the traveling wave $\phi_\text{a} = \frac{1}{2} \cos{\frac{x- x_0}{\eta} } + \frac{1}{2}$
was fitted with respect to $x_0$ to the numerical solution at several time points.
Then the mean squeared deviation between numerical fitted analytical form $\phi_\text{a}(x_i)$ and the actual numerical values $\phi^m_i$ was calculated
\begin{align}
\text{MSD} = \frac{1}{N} \sum_{i=1}^{N} \Bigl( \phi_\text{a}(x_i) -  \phi^m_i \Bigr) ^2 
\end{align}
with $x_1, ..., x_N$ belonging to the diffuse interface.
The MSD-value is subject to fluctuations in time, but nevertheless it has well defined time averaged values $\overline{\text{MSD}}$, around which fluctuations take place and which deliver a good measure for the reproduction quality of the traveling wave solution.
In order to see the correlation between $\overline{\text{MSD}}$ and the relative error in velocity $(v-v_{th})/v_{th}$, the respective values are plotted together in figure~\ref{fig_CorrPlot}.
Obviously the error in kinetics depends on the error in the traveling wave reproduction, and with reducing the latter one the first one can reduced without higher numerical effort, which is a good argument for the usage of constant stabilization methods.
On top of that it can be seen, that the numerical effects are causing greater errors than the changes in interface peclet numbers.
Note that for a given factor $f$, given numerical method and given spatial resolution the respective points in figure~\ref{fig_CorrPlot} differ only by their peclet number. 
Those points are clustering closely to each other.

It should be mentioned, that the constant stabilization methods exhibit artificial behaviour for very small velocities. They tend to trap the traveling wave solution in an artificial stationary solution, where the stabilization operator compensates in every gridpoint the driving force term. However for intermediate velocities this problem doesn't occur.

\subsection{Comparison between local and non local mobility correction}
\begin{figure}
\centering
\includegraphics[width = 0.75\linewidth]{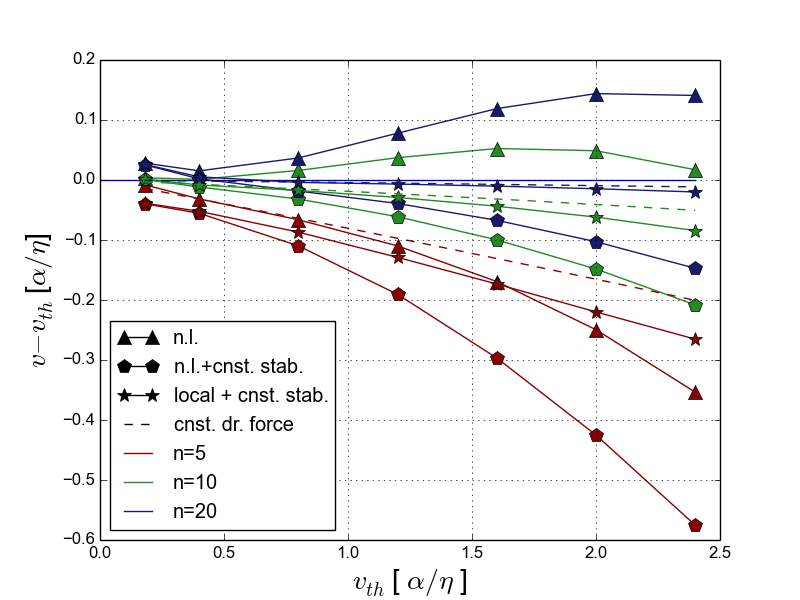}
\caption{Velocities of the traveling wave solution for different discretizations $n=\eta/\Delta x$ from simulations with local mobility correction (stars), non local mobility correction (pentagons and triangles), and simulations with a spatially constant driving force imposed (dashed lines). 
Simulations with non local mobility correction were performed with a constant stabilization operator (pentagons)  and without (triangles).
Different physical mobilities $M^\text{phys}$ were chosen to be reproduced resulting in different expected interface velocities $v_\text{th}$, which are shown on the horizontal axis. 
All units are scaled such, that the numerical value represents the corresponding interface peclet number $p_e$. 
On the y-axis the deviation of the actual numerical velocities from the theoretically expected velocity $v_\text{th}$ is plotted. }
\label{fig_Mvar_nvar}
\end{figure}

By comparing results for different spatial resolutions $n=\frac{\eta}{dx}$ in figure~\ref{fig_Mvar_nvar} one important finding can be made:
The local mobility correction scheme works almost equally precise as the non local one for poor spatial resolutions like $n=5$.
In order to gain an advantage in the precision of kinetics good spatial resolutions like $n=10$ or higher are required.
This can be well understood, since we cannot expect from a theory enhanced by local information to deliver better results in simulations, when the local information is numericaly reduced to a minimum.

Moreover this is together with the results in figure \ref{fig_CorrPlot} a clear indicator, that numerical errors have a stronger impact than errors arising from the diffusivity of the interface.
Nevertheless the results obtained with local mobility correction behave more predictable with staying below the constant driving force results, whereas the non local results are strongly influenced by the applied numerical method and discretization in wether their velocities over- or undershoot.

Additionally the local mobility correction scheme is giving more accuracy compared to the non local one for higher peclet numbers.
Even for low spatial resolution this effect can be seen.
This is expected, because for higher peclet numbers the diffuse interface errors are getting more severe.

\section{Conclusion}

In this work we developed an extended approach to thin interface limit analyses, which includes local information in the mobility factor and aims to make quantitative agreement between phase-field and sharp interface models possible with higher accuracy for higher peclet numbers.
Since the common thin interface limits already mitigate diffuse interface errors very reliably, most of quantitative errors in phase-field simulations are of numerical nature.
Consequently our local approach grants significant advantage compared to the common method only under specific conditions: high spatial resolution and moderately high interface peclet numbers.
Moreover the local mobility correction scheme as presented here, is applicable for 1D problems and not trivially extendes to 2D or 3D situations, where interface velocities and diffusion lengths can vary along the interface.
In the light of recent research, that aims to remove numerical errors in phase-field simulations \cite{Finel2018}, the local mobility correction might deliver additional benefit in combination with mitigation measures for numerical errors.


\section*{Acknowledgments}
This work was funded by the Deutsche Forschungsgemeinschaft (DFG) under grants no. Ste1116/10-2.

Declaration of interest - None. 

\clearpage



\bibliography{citations.bib}

\bibliographystyle{elsarticle-num}

\newpage

\end{document}